\documentclass[aps,prx,twocolumn,showpacs,superscriptaddress,groupedaddress]{revtex4}  
\usepackage{graphicx}  
\usepackage{dcolumn}   
\usepackage{bm}        
\usepackage{amssymb}   
\usepackage{color}
\usepackage{ulem}
\usepackage[utf8]{inputenc}
\usepackage{amsmath}
\usepackage{hyperref}
\usepackage{mathtools}
\usepackage{soul}
\hypersetup{
    colorlinks=true,
    linkcolor=blue,
    filecolor=magenta,      
    urlcolor=cyan,
}

\begin{document}



\title{Coarse-grained entropy production with multiple reservoirs: unraveling the role of time-scales and detailed balance in biology-inspired systems}
\author{Daniel M. Busiello}
\affiliation{Ecole Polytechnique F\'ed\'erale de Lausanne (EPFL), Institute of Physics Laboratory of Statistical Biophysics, 1015 Lausanne, Switzerland}
\author{Deepak Gupta}
\affiliation{Dipartimento di Fisica `G. Galilei', INFN, Universit\'a di Padova, Via Marzolo 8, 35131 Padova, Italy}
\author{Amos Maritan}
\affiliation{Dipartimento di Fisica `G. Galilei', INFN, Universit\'a di Padova, Via Marzolo 8, 35131 Padova, Italy}
\date{\today}

\begin{abstract}
A general framework to describe a vast majority of biology-inspired systems is to model them as stochastic processes in which multiple couplings are in play at the same time. Molecular motors, chemical reaction networks, catalytic enzymes, and particles exchanging heat with different baths, constitute some interesting examples of such a modelization. Moreover, they usually operate out of equilibrium, being characterized by a net production of entropy, which entails a constrained efficiency. Hitherto, in order to investigate multiple processes simultaneously driving a system, all theoretical approaches deal with them independently, at a coarse-grained level, or employing a separation of time-scales. Here, we explicitly take in consideration the interplay among time-scales of different processes, and whether or not their own evolution eventually relaxes toward an equilibrium state in a given sub-space. We propose a general framework for multiple coupling, from which the well-known formulas for the entropy production can be derived, depending on the available information about each single process. Furthermore, when one of the processes does not equilibrate in its sub-space, even if much faster than all the others, it introduces a finite correction to the entropy production. We employ our framework in various simple and pedagogical examples, for which such a corrective term can be related to a typical scaling of physical quantities in play.
\end{abstract}

\maketitle

\section{General models for multiple coupling}
\label{sec:intro}

Biological systems in general operate out of equilibrium \cite{prigogine}. These can be described in terms of different states (both discrete and continuous), which are connected to each other through a set of transitions with given rates. States of a system can be of various kinds, e.g. these can represent different chemical species or configurations \cite{semenov,rao}, as well as the coupling to a given bath or a given potential \cite{schn,diffbaths}, just to cite some examples.

\begin{figure*}
  \begin{center}
    \includegraphics[width=12cm]{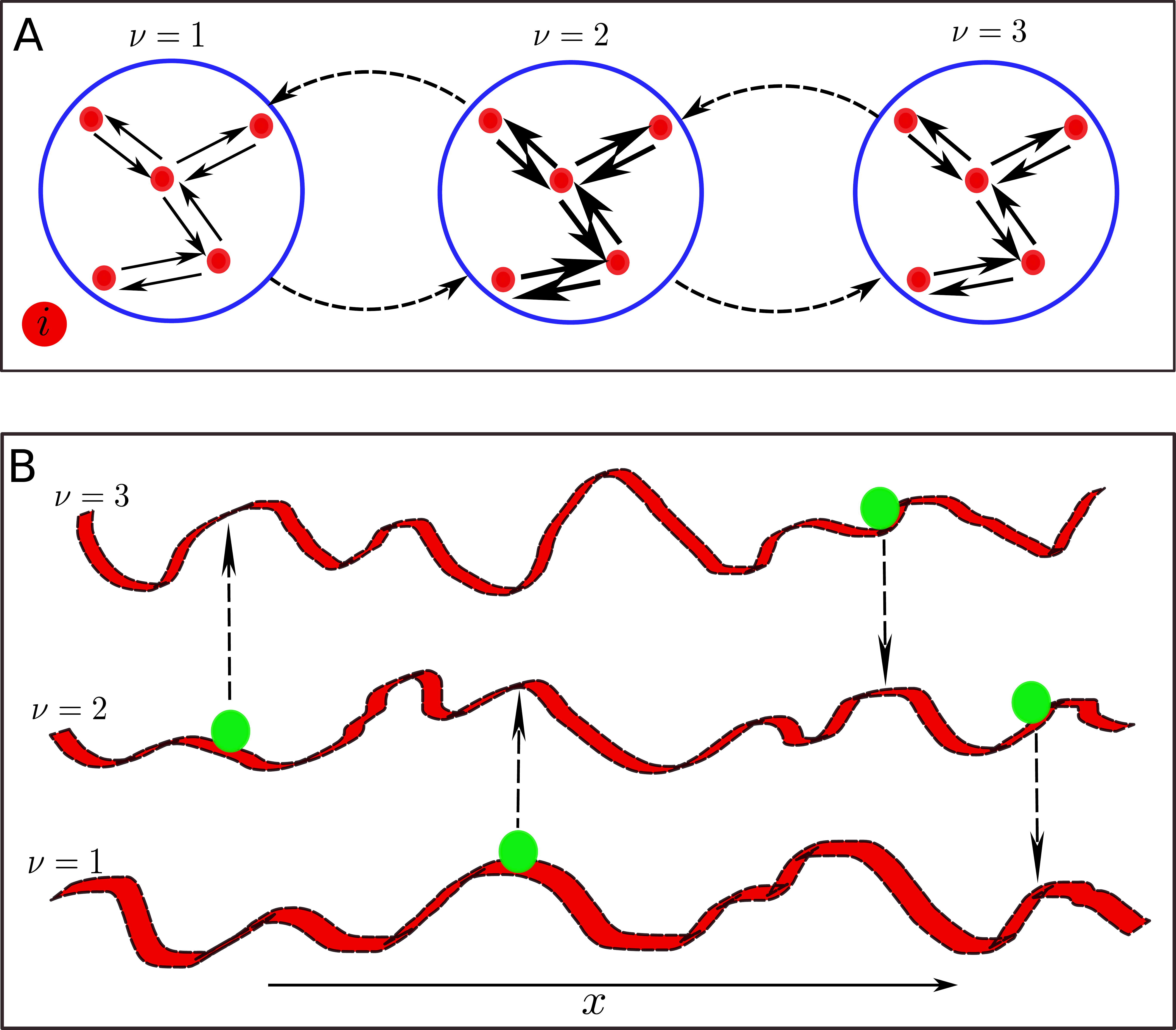}
    \caption{Systems with multiple coupling. A) a class of transitions is identified by a fixed $\nu$ and a change in the index $i$ (within the blue circle). The other class of transitions is associated to a variation of $\nu$ for a fixed $i$ (black dashed lines). B) One degree of freedom is continuous, representing the position in space $x$. Some transitions changes $\nu$ for a fixed $x$ (black dashed lines), while some others move the particles in the same red stripe, varying $x$ while keeping $\nu$ fixed.}
    \label{fig:system}
  \end{center}
\end{figure*}

In general, multiple processes can act on a system at the same time, each one being responsible for transitions between states of the same kind. As an example, chemical species that can also diffuse in space are connected though chemical reactions \cite{diss-sel,fickslaw}, while the diffusive mechanism is governed by the Fick's law. Alternatively, particles diffusing in a solution that can be connected to different baths follow a Fokker-Planck equation \cite{gardiner,Schmiedl_2007,PhysRevLett.99.230602}, while the switching between baths is controlled by a different process. In Fig. \ref{fig:system} we present two examples of systems with multiple coupling.

A system composed only of discrete states is shown in Fig. \ref{fig:system}A. It can perform transitions in different sub-spaces: within each single circle, identified by an index $\nu$, and from one circle to another, changing $\nu$, in an abstract reservoir-space. This scheme fits the modelization of the motion of bio-molecules switching among baths at different temperatures \cite{diss-sel}, such as proteins with many configurations \cite{exc-hyd}, or chemical species interacting with the solution in which they are embedded \cite{Wachtel_2018}.


Some degrees of freedom can also be continuous, e.g.  a position in space. The sketch in Fig. \ref{fig:system}B represents this situation. Molecular motors, in which each red line corresponds to a track \cite{parmeggiani1999energy,julicher}, or diffusing enzymes, where to different $\nu$ a different diffusion coefficient (and a different chemical state) is associated \cite{sen}, are  clearly examples belonging to this class of systems.

From a theoretical point of view, a complete framework to model a system in presence of multiple coupling is provided by a Master Equation keeping track of each process in play \cite{gardiner}. As discussed above, several biology-inspired models can be constructed within this picture. In what follows, they will serve mainly as inspiration for the study of more fundamental and general aspects.

We start with a system composed of discrete states only, as in Fig. \ref{fig:system}A. Each state is characterized by two indices, $i$ and $\nu$, which label the accessible sub-spaces, named $i$- and $\nu$-space for the sake of simplicity. Hence, the probability to be in the state $(i,\nu)$ is $p_i^\nu$ with $i=1,\dots,N$ and $\nu=1,\dots,n$. 
The evolution equation for $p_i^\nu$ is \cite{van1992stochastic}
\begin{equation}
\frac{dp_i^\nu(t)}{dt} = \left( \boldsymbol{W}^\nu \vec{p}^\nu(t) \right)_i + \left( \boldsymbol{\Phi}_i \vec{p}_i(t) \right)^\nu,
\label{MEmult}
\end{equation}
where $\boldsymbol{W}^\nu$ is the $N\times N$ transition matrix with off-diagonal elements $(ij)$ equal to $w_{j \to i}^\nu$, i.e., the transition rate from state $(j, \nu)$ to state $(i, \nu)$. Analogously, $\boldsymbol{\Phi}_i$ is the $n\times n$ transition matrix with off-diagonal elements $(\nu, \mu)$ equal to $\phi^{\mu \to \nu}_i$, i.e., the transition rate from state $(i, \mu)$ to state $(i, \nu)$. In this short notation, $\vec{p}^\nu = (p^\nu_1,... p^\nu_N)^\top$ and $\vec{p}_i = (p_i^1,... p_i^n)^\top$, where $\top$ refers to the transpose operator. Thus, for example, $\left( \boldsymbol{W}^\nu \vec{p}^\nu(t) \right)_i\equiv \sum_j (w_{j \to i}^\nu p_j^\nu-w_{i \to j}^\nu p_i^\nu)$.The diagonal elements of the two matrices are given by $(\boldsymbol{W}^\nu)_{ii}= -\sum_j w_{i \to j}^\nu $ and $(\boldsymbol{\Phi}_i)_{\nu \nu}=-\sum_\mu \phi^{\nu \to \mu}_i$. This choice guarantees that the probability distribution $p_i^\nu$ is normalized at all times: $\sum_{i=1}^N \sum_{\nu=1}^n p_i^\nu(t)=1$. 

In general, the transition rates acting on the index $i$ can also depend on $\nu$, and the ones governing the transitions in the $\nu$-space, can also change with $i$. This is the case, for example, of chemical rates between species $i$, depending on the temperature of the bath $\nu$ which the system is coupled to \cite{diss-sel}.


When one degree of freedom is continuous, it is immediate to rephrase the above formalism into a differential equation considering contributions from the dynamics on the discrete set of variables as well as on the continuous ones. Therefore, the evolution of the probability to be in the state $(x,\nu)$ at time $t$, $P^\nu(x,t)$. is \cite{gardiner,van1992stochastic}: 
\begin{equation}
\partial_t P^\nu(x,t) = -\partial_x J^\nu(x,t) + \left( \boldsymbol{\Phi}(x) \vec{P}(x,t) \right)^\nu.
\label{FPEmult}
\end{equation}
Here, for the sake of simplicity, we restrict ourselves to one-dimension spatial systems, for which analogies and differences with respect to a description in terms of discrete states is well-known and extensively studied \cite{busielloENT,jstat,seifFT,esposito-CG}. Nonetheless, the generalization to higher dimensions is straightforward. In the above equation, $J^\nu(x,t)$ is the probability flux at position $x$ and time $t$, that can also depend on $\nu$. The detailed structure of the probability flux will be discussed later. Molecular motors are the most prominent examples in this category, where the potential experienced by the motor, encoded in $J^\nu(x,t)$, depends on the track $\nu$ on which it is moving \cite{julicher}.

Notice that we are implicitly assuming that the system can either undergo a transition in the $\nu$ space, for a fixed $i$ (or $x$), or it can change the label $i$ to $j$ (or going from $x$ to $x+dx$) remaining in the same state $\nu$. Indeed, this is a reasonable assumption if a suitable time-scale exists over which only one transition at a time can occur \cite{gillespie}. However, we leave for future works the investigation of cases where such a time-scale does not exist and thus also processes where both indices $i$ (or $(x)$) and $\nu$ are allowed to change in a single transition. Moreover, even though the system described above can be mimicked by time-periodic rates in the absence of multiple coupling \cite{busiello-raz}, we will not deal with the latter picture herein. 

In what follows, without loss of generality, we will consider the dynamics in the $\nu$-space to be faster than the one in the $i$-space, unless otherwise stated.

Since these models allow for a complete description of systems out-of-equilibrium, the main focus of this work is to study the net production of entropy in the surroundings, which is one of the fingerprints of a non-equilibrium condition \cite{busielloENT,pigo,seifEP,unidirectional}. Besides its paramount theoretical importance, recently, the entropy production is getting much attention also from an experimental perspective. Indeed, being this involved in the celebrated uncertainty relations \cite{baratotur,pietztur,gingrichtur,falascotur,barato,gupta2020thermodynamic,friedman2020thermodynamic}, through them it might be useful to infer the dissipation in a biological system, hence quantifying how far from equilibrium they are operating \cite{horo,hyper,guptainfer,otsubo2020estimating,van2020entropy}. It has also been shown that this quantity play a leading role in driving the selection process in a chemical reaction network, being able to estimate how much thermal energy is converted into chemical one \cite{diss-sel}. Recently, in \cite{metab}, an upper bound to Gibbs energy dissipation rate is found to constrain intracellular metabolic fluxes. The entropy production is also a key quantity to estimate the efficiency of non-equilibrium machines \cite{qian,SKM}, and to eventually build artificial motors with a performance as close as possible to natural ones \cite{jarz-raz,busiello-raz}.
However, several other observables (e.g. heat, spatial currents) may spark intriguing questions in the fields of bioenergetics and non-equilibrium thermodynamics, and we believe they deserve a detailed investigation in future works.

The entropy production of a system with multiple coupling is a problem that has been faced several times from various perspectives \cite{diss-sel,qian,esposito-CG,diffbaths}. The most general approach is to evaluate the entropy production by considering all processes acting on similar time-scales \cite{mehl,kawa,uhl,eff-th,pep-1,pep-2,pep-3}. In this case, the result is devoid of approximations. However, this is not always the case. In many situations, the exact rates characterizing all the processes are not known, and some simplifications have to be employed. 

In this paper, we consider two different models depicted in Fig. \ref{fig:system} mimicking physical systems as discussed above, Eqs. \eqref{MEmult} and \eqref{FPEmult}. We aim at evaluating the entropy production for such systems driven by multiple processes at the same time, when some of them are faster than the others. In the presence of a time-scale separation, one would naively think that the system is evolving under an effective dynamics, with its energetics directly derived from the latter. Instead, we present a general theory to consider the various possible approximations due to relative temporal-scale, leading, in general, to different results for the entropy production. Well-known formulas presented in literature as general results emerge from our framework only under some limiting conditions.

\section{Outline}
\label{outline}
In the following, we briefly present the outline of the paper before getting into the detailed discussion on the various forms of the entropy productions here presented. All possible approximations stemming from our framework fall into two classes, depending on whether the time-scale separation is performed before or after the evaluation of the entropy production.

The first one, which we refer to as \textit{Coarse Grained Approximation} (CGA), deals with the coarse-graining of the dynamics \eqref{MEmult} or \eqref{FPEmult}, by integrating over all the possible fast states (see Sec. \ref{sec-cga}), \textit{before} the quantification of the entropy production. Herein, we consider the dynamics in the $\nu$-space to be relatively faster than the internal one over $i$ variables. We show that the total entropy produced using such coarse-grained dynamics contains only \textit{effective} probability distributions and transition rates.

In the second approach [\textit{Single Index Approximation} (SIA)], the total entropy production (see Sec. \ref{sec:single}) carries details of fast and slow processes, through their transition matrices, $\textbf{W}$ and $\boldsymbol{\Phi}$. However, the time-scale separation is employed \textit{afterwards} to simplify the expression of the probability distributions. It implies that the dynamics over the fast space, i.e., the $\nu$-space in this paper, is always at stationarity. This approach, under further approximations leads to an expression for the entropy production given in \cite{esposito-CG,diffbaths} for the case of a system in contact to several baths, one at a time.

Notice that, while in SIA the information about fast states is neglected only in the probabilities, in CGA it is ignored also in the rates, replacing them with effective quantities. In other words, CGA is much stronger than SIA, in the sense that the latter is aware of some microscopic details that are neglected using the former approximation. As a consequence, the entropy production obtained through CGA is always less than or equal to that derived using SIA (see Sec. \ref{sec:single}). 

Consider an experimental setting. When fast processes cannot be observed, the only solution to compute the entropy production is to employ CGA. However, it is possible that fast transition rates are known from different experiments for every possible fixed
realization of the variable $\nu$, whereas the occupancy of each fast state cannot be measured (e.g. fluorescence microscopy for enzymes \cite{granick, chemo}). In this case, SIA can be used to obtain a better estimate for the entropy production with respect to CGA. Moreover, it is true that SIA can also be employed to speed up the numerical evaluation of the entropy production in systems with a particularly large state-space.

The two models here studied, Eqs. \eqref{MEmult} and \eqref{FPEmult}, are manifestly very general, since they contain no approximation on the dynamics, and therefore, can capture phenomena not encoded in the  above mentioned simplified descriptions, i.e., in CGA and SIA. Here, we show that the presence of the degrees of freedom of the $\nu$-space introduces two novel ingredients that have to be taken into account. The first one is a characteristic time-scale, considering the relative fast dynamics over the $\nu$-space. The other one is whether or not both dynamics asymptotically drive the system towards an equilibrium state in their respective subspace. 

We start by formally deriving the entropy production in the general case and in both the approximation schemes detailed above. In the limit in which the fast processes are detailed balanced (see Secs. \ref{sec:single} and \ref{SI-GB}), i.e., these would drive the system toward equilibrium in their subspace, we will obtain well-known results present in the literature \cite{esposito-CG,esposito-3f,diffbaths}. The advantage of our approach is that we can relate the emerging \textit{effective} quantities to the ones characterizing the microscopic complete picture.

Further, when the detailed balance is  broken, the formula for the entropy production is affected by the interplay between non-equilibrium features of the fastest process and its characteristic time-scale. Said differently, out-of-equilibrium conditions generate interactions that couple time-scales that would be separated otherwise. We also show that, when the system is close to equilibrium at stationarity (see Sec. \ref{BDB}), a scaling relation holds determining whether such an interplay is relevant or not for the quantification of the entropy production.



In details, the remaining paper is organized as follows. In Sec.~\ref{sec:one}, we discuss the entropy production for a system with both discrete $i$- and $\nu$-space. Subsec. \ref{ts-d-d} presents the time-scale separation procedure. CGA and SIA are presented and discussed in Subsecs. \ref{sec-cga} and \ref{sec:single}, respectively. Subsec. \ref{SI-GB} refers to the simplest case of $i$-independent transitions in the $\nu$-space. Further, the correction to the entropy production is obtained in Sec. \ref{BDB} when the transitions in the $\nu$-space are the fastest, but their rates do not satisfy detailed balance. Finally, we give some examples to illustrate our results in Sec. \ref{examples}, evidencing the physical meaning of the condition for a non-vanishing correction to the entropy production due to non-equilibrium features in some simple pedagogical models. We conclude our paper in Sec. \ref{conc}. In the appendix, we present the detailed discussion on systems which make transitions among both discrete and continuous states, following the same structure exploited in the main text. 


\section{Entropy production with transitions among discrete states}
\label{sec:one}

In Sec. \ref{sec:intro}, we introduced two different models for system with multiple coupling, Eqs. \eqref{MEmult} and \eqref{FPEmult}. Here, we first consider a system that performs jumps in the discrete $i$-space, as well as in the $\nu$-space. Its evolution equation is thus given by Eq. \eqref{MEmult}. 

The (average) entropy of the system is given by \cite{schn}
\begin{align}
S_{sys}:=\sum_\nu S^\nu_{sys},
\label{Ssys}
\end{align}
where
\begin{align}
S^\nu_{sys}=-\sum_{i=1}^{N}p_i^\nu\log p_i^\nu,
\end{align} 
is the entropy of the system when it is only coupled to a single state $\nu$. The sum in the last equation is performed over the state variable $i$. The total system entropy production is obtained by differentiating the Eq. \eqref{Ssys} with respect to time: 
\begin{eqnarray}
\dot S_{sys} & = & \sum_{\nu=1}^n\dot S^\nu_{sys}=-\sum_{\nu=1}^n\sum_{i=1}^{N}\big[\dot p_i^\nu\log p_i^\nu+\dot p_i^\nu\big] \nonumber \\
& = & -\sum_{\nu=1}^n\sum_{i=1}^{N}\dot p_i^\nu\log p_i^\nu ,
\label{ent-mult}
\end{eqnarray}
where  $\sum_{i,\nu} \dot p^\nu_i=0$ has been used in the last step, due the probability conservation.
Proceeding like in Ref. \cite{schn}, we can re-write the total system entropy production as 
\begin{align}
\dot{S}_{sys}=\overbrace{\dot{S}_{tot}+\dot{S}^{X}_{tot}}^{\dot{\mathcal{S}}_{tot}}-\overbrace{(\dot{S}_{env}+\dot{S}^{X}_{env})}^{\dot{\mathcal{S}}_{env}},
\label{S-mult}
\end{align}
where subscripts $sys$, $env$, $tot$, refer to system, environment, and total, respectively. The superscript $X$ only indicates the $\nu$-space. In the above equation, we identify the terms as follows
\begin{align}
\label{stotmult}
\dot S_{tot}&=\sum_{\nu}\sum_{i,j} w^\nu_{j\to i}p^\nu_j \log\dfrac{w^\nu_{j\to i}p^\nu_j}{w^\nu_{i\to j}p^\nu_i},\\
\dot S_{env}&=\sum_{\nu}\sum_{i,j} w^\nu_{j\to i}p^\nu_j \log\dfrac{w^\nu_{j\to i}}{w^\nu_{i\to j}},\\
\dot S^{X}_{tot}&=\sum_{i}\sum_{\mu,\nu}\phi_i^{\mu\to \nu}p^\mu_i\log \frac{\phi_i^{\mu\to\nu}p_i^\mu}{\phi_i^{\nu\to\mu}p_i^\nu}, \label{sxenv-eq}\\
\dot S^{X}_{env}&=\sum_{i}\sum_{\mu,\nu}\phi_i^{\mu\to \nu}p^\mu_i\log \frac{\phi_i^{\mu\to\nu}}{\phi_i^{\nu\to\mu}},\label{sxenv-eq}
\end{align}
where, $\dot S^X \equiv \dot{S}^{X}_{tot}-\dot{S}^{X}_{env}$ is the system entropy production due to the process that governs transitions between different states belonging to the $\nu$-space. 

In Eq. \eqref{S-mult}, $\dot{\mathcal{S}}_{tot}$ and $\dot{\mathcal{S}}_{env}$, respectively, are the total entropy production and the environmental entropy production due to both transitions within $i$- and $\nu$-space. In Eqs. \eqref{stotmult}--\eqref{sxenv-eq} we have separated the contribution to the entropy production given by the transition matrix $\boldsymbol{W}^\nu$, which couples states in the $i$-space for each state $\nu$, from the one given by $\boldsymbol{\Phi}_i$, acting on the index $\nu$ for a given $i$. So far, we have not used any approximation, thus the entropy production we have derived contains all the available information about the system.

Similarly, the entropy production for the system obeying Eq. \eqref{FPEmult} is given in  Appendix \ref{ep-c-d} [see Eq. \eqref{SdotFPE}].

In the following, we analyse CGA and SIA for the dynamics \eqref{MEmult}. To do so, we start with introducing the time-scale separation procedure, which is a fundamental ingredient for both approximations.
\\

\subsection{Time-scale separation on the dynamics}
\label{ts-d-d}

Let us first consider a system with $N$ states, amenable to be described by a Master Equation governed by the transition matrix $\boldsymbol{W}^\nu$, and coupled to $n$ states in the $\nu$-space. The whole dynamical evolution is described as in Eq.~\eqref{MEmult}. From a physical perspective, states in the $\nu$-space may correspond to reservoirs of thermal energy, matter, and so on \cite{schn}, each of them driving the system away from equilibrium.

Just to fix some ideas, we provide one illustrative example to qualitatively understand the possible scenarios. A molecule has different states: it can change its configuration, or interact with the solution forming complexes or varying its chemical composition. Each state is identified by a certain $i = 1,\dots ,N$. Moreover, the molecule can be coupled to several thermal baths, one at a time, each one identified by an index $\nu = 1,\dots,n$ (black dashed lines in Fig. \ref{fig:system}A). Hence, the bath temperature modifies the chemical rates $(\boldsymbol{W}^\nu)$, and the molecular state can influence, in turn, the switching between baths $(\boldsymbol{\Phi}^i)$ (e.g.  employing a positive feedback for a chemical selection \cite{diss-sel}). Three possibilities have to be considered: 
\begin{itemize}
    \item[i)] chemical reactions affecting the state of the molecule eventually would lead the system to a non-equilibrium condition, if the index $\nu$ were fixed. Mathematically speaking, the matrix $\boldsymbol{W}^{\nu}$ is not detailed balanced. On the contrary, the switching between baths is unbiased, so that, for a fixed $i$, $\boldsymbol{\Phi}_i$ is detailed balanced. In other words, the stationary probability distribution in the $\nu$-space only, for a fixed $i$, $\pi_i^\nu$ is such that $\pi^\mu_i \phi^{\mu \to \nu}_{i} = \pi^\nu_i \phi^{\nu \to \mu}_{i}$;
    \item[ii)] $\boldsymbol{W}^{\nu}$ is detailed balanced whereas $\boldsymbol{\Phi}_i$ is not, meaning that the switching process among several baths drives the system out of equilibrium;
    \item[iii)] both transition matrices are not detailed balanced, and an interplay between the two time-scales characterizing the processes can lead to non-trivial situations.
\end{itemize}


Here, we aim at investigating how the expression of the entropy production changes when the transitions taking place in a given subspace (herein, $\nu$-space) are relatively faster than those occurring in the other (herein, $i$-space).  To do so, we introduce below the standard framework to employ a time-scale separation in the dynamics. We also show, in the next Sections, how to construct CGA and SIA. In particular, we show that in cases analogous to i) the SIA leads to a well-known formula reported in the literature \cite{esposito-CG,diffbaths}, while in cases belonging to the class ii) and iii) additional terms arise due to the interplay between non-equilibrium stationarity and the time-scale of the fastest process.

Similar analysis for a system diffusing along a one-dimensional domain and with fast transitions in the $\nu$-space, Eq. \eqref{FPEmult}, is described in detail in Appendix \ref{ts-cd-model}.

We introduce a characteristic scale, $1/\epsilon$ and $0<\epsilon \ll1$, such that the matrix element $[\boldsymbol{\Phi}]^{\mu\to\nu} \to \epsilon^{-1}[\tilde{\boldsymbol{\Phi}}]^{\mu\to\nu}$, and the Master Equation \eqref{MEmult} becomes:
\begin{align}
\dfrac{dp^\nu_i}{dt}&=\sum_{j=1}^{N} (w^\nu_{j\to i}p^\nu_j-w^\nu_{i\to j}p^\nu_i)~+\nonumber\\
& + \epsilon^{-1} \sum_{\mu=1}^{n} (\tilde{\phi}^{\mu\to \nu}_ip^\mu_i-\tilde{\phi}^{\nu\to \mu}_ip^\nu_i),
\label{mult}
\end{align}

In order to solve the system, we assume the solution of the above differential equation to be 
\begin{equation}
p^\nu_i = p^{\nu 0}_i + \epsilon^\beta p^{\nu 1}_i +\epsilon^{2\beta} p^{\nu 2}_i + \text{higher orders},
\label{expansion}
\end{equation}
with the constant $\beta > 0$. Since $\sum_{i,\nu}p^\nu_i=1$ we must have
\begin{align}
& \sum_{i,\nu}p^{\nu 0}_i=1, \\
& \sum_{i,\nu}p^{\nu k}_i=0, \quad \forall\quad k>0.
\label{normcond}
\end{align}
Inserting the above solution in Eq. \eqref{mult}, we obtain 
\begin{align}
\dfrac{dp_i^{\nu 0}}{dt} + \epsilon^\beta \dfrac{dp_i^{\nu 1}}{dt} +\cdots &= \sum_j(w^\nu_{j\to i} p^{\nu 0}_j-w^\nu_{i\to j} p^{\nu 0}_i) +\nonumber\\
& + ~\epsilon^\beta \sum_j(w^\nu_{j\to i} p^{\nu 1}_j-w^\nu_{i\to j} p^{\nu 1}_i) + \nonumber \\
& + ~\epsilon^{-1} \sum_{\mu=1}^{n} (\tilde{\phi}^{\mu\to \nu}_ip^{\mu 0}_i-\tilde{\phi}^{\nu\to \mu}_ip^{\nu 0}_i)+\nonumber\\
& + ~\epsilon^{\beta-1} \sum_{\mu=1}^{n} (\tilde{\phi}^{\mu\to \nu}_ip^{\mu 1}_i-\tilde{\phi}^{\nu\to \mu}_ip^{\nu 1}_i)+\nonumber\\
& + ~\epsilon^{2\beta-1} \sum_{\mu=1}^{n} (\tilde{\phi}^{\mu\to \nu}_ip^{\mu 2}_i-\tilde{\phi}^{\nu\to \mu}_ip^{\nu 2}_i) .
\end{align}
Let us first consider the case when $\beta=1$. Here, we equate terms of the same order in $\epsilon$ on both sides, finding:
\begin{align}
0&=\sum_{\mu=1}^{n} (\tilde{\phi}^{\mu\to \nu}_ip^{\mu 0}_i-\tilde{\phi}^{\nu\to \mu}_ip^{\nu 0}_i)\label{solveeq}\\
\dfrac{dp_i^{\nu 0}}{dt}&= \sum_j(w^\nu_{j\to i} p^{\nu 0}_j-w^\nu_{i\to j} p^{\nu 0}_i)+\nonumber\\
& +\sum_{\mu=1}^{n} (\tilde{\phi}^{\mu\to \nu}_ip^{\mu 1}_i-\tilde{\phi}^{\nu\to \mu}_i p^{\nu 1}_i),\label{pn0}\\
\dfrac{dp_i^{\nu 1}}{dt} &=\sum_j (w^\nu_{j\to i} p^{\nu 1}_i-w^\nu_{i\to j} p^{\nu 1}_j) \nonumber\\
& +\sum_{\mu=1}^{n} (\tilde{\phi}^{\mu\to \nu}_ip^{\mu 2}_i-\tilde{\phi}^{\nu\to \mu}_ip^{\nu 2}_i)\label{pn1}.
\end{align}
The elements $p^{\nu 0}_i$ always satisfy Eq. \eqref{solveeq}, which implies the stationarity of the zeroth order of the probability density function $p_i^\nu$, for each $i$, with respect to the dynamics of the $\nu$-space. Intuitively, the system reaches stationarity in the fastest space before performing a transition in the slow one.

Conversely, if we equate terms of same order in $\epsilon$ in the case of $\beta\neq 1$, we obtain the following equations
\begin{eqnarray}
0&=&\sum_{\mu=1}^{n} (\tilde{\phi}_i^{\mu\to \nu}p^{\mu 1}_i-\tilde{\phi}_i^{\nu\to \mu}p^{\nu 1}_i),\\
0&=&\sum_{\mu=1}^{n} (\tilde{\phi}_i^{\mu\to \nu}p^{\mu 0}_i-\tilde{\phi}_i^{\nu\to \mu}p^{\nu 0}_i).
\end{eqnarray}
suggesting that $p_i^{\nu 0} \propto p_i^{\nu 1}$. Therefore, we must have $\beta=1$, as in the standard approach \cite{gardiner}.

Then, to the order $\epsilon^{-1}$, solving Eq.~\eqref{solveeq}, we can write the zeroth order solution as 
\begin{equation}\label{0thorder}
p^{\nu 0}_i = p_i \pi^{\nu}_i, 
\end{equation}
such that the stationary probability distribution for the $(\boldsymbol{\Phi}^i)$ matrix, $\pi^\nu_i$ for each $i$ in the $\nu$-space is normalized, i.e., $\sum_\nu \pi^\nu_i = 1$. 

Substituting the zeroth order solution $p_i^{\nu 0}$ in  Eq.~\eqref{pn0} and summing over the fast states $\nu$, we obtain the evolution equation for $p_i$ as 
\begin{equation}
\dfrac{dp_i}{dt} = \sum_j \big[ \tilde{w}_{j\to i} p_j - \tilde{w}_{i\to j} p_i \big].
\label{solp}
\end{equation}
where we have defined the effective transition rates 
\begin{equation}\label{weff}
\tilde{w}_{j\to i}:= \sum_\nu \pi^\nu_j w^\nu_{j\to i}
\end{equation}
Notice that after Eq.~\eqref{solp} is solved, with the appropriate initial conditions, $p_i^{\nu 0}=p_i\pi^\nu_i$ is determined using Eq. \eqref{0thorder}. Hence, Eqs. \eqref{normcond}, \eqref{pn0}, and the equation obtained by summing over $\nu$ Eq.~\eqref{pn1} can be used to determine $p_i^{\nu 1}$. The higher order correction to Eq. \eqref{expansion} can thus be calculated iteratively.

Thus, we have an evolution described in terms of coarse-grained rates, which are nothing but ensemble averages of the transition rates, $w^\nu_{j\to i}$, and coarse-grained probabilities $p_i$. In some experimental situations, we might think to them as the only \textit{accessible} variables. When this is the case, CGA has to be employed, leading to an entropy production which depends solely on these variables (see next subsection).

\subsection{Integrating the fastest states (CGA)}
\label{sec-cga}
In terms of the coarse-grained probabilities $p_i$'s, the system entropy production is defined as:
\begin{equation}\label{SGE}
S_{sys}(p) = - \sum_{i} p_i \log p_i.
\end{equation}
Differentiating both sides with respect to time, and using Eq. \eqref{solp} we get:
\begin{equation}
\dot{S}_{sys}(\tilde{w},p)=\dfrac{1}{2}\sum_{i,j} (\tilde{w}_{i \to j} p_i - \tilde{w}_{j \to i} p_j) \log \frac{p_i}{p_j}.
\end{equation}
We can now define the corresponding environmental contribution, in terms of the coarse-grained variables:
\begin{equation}\label{env-new}
\dot{S}_{env}(\tilde{w},p)=\dfrac{1}{2}\sum_{i,j} (\tilde{w}_{i \to j} p_i - \tilde{w}_{j \to i} p_j) \log \frac{\tilde{w}_{i \to j}}{\tilde{w}_{j \to i}},
\end{equation}
 so that the total entropy production in this coarse-grained description becomes
\begin{align}
\dot{S}_{tot}(\tilde{w},p)&= \dot{S}_{sys}(\tilde{w},p)+\dot{S}_{env}(\tilde{w},p) \nonumber \\
&= \dfrac{1}{2}\sum_{i,j} (\tilde{w}_{i \to j} p_i - \tilde{w}_{j \to i} p_j) \log \frac{\tilde{w}_{i \to j} p_i}{\tilde{w}_{j \to i} p_j}.
\label{Seff}
\end{align}

The environmental entropy production in Eq. \eqref{env-new} is different from the one we introduced before, i.e., in the most general case [Eq.~\eqref{S-mult}], since it depends solely on variables we consider as observables of the system.

In the case of a system with continuous variables (for example, the system shown in Fig.~\ref{fig:system}B), the coarse-graining procedure on the fast space yields the entropy production given in Eq.~\eqref{cg-c-d-sys}, where again details of the information on the dynamics in the $\nu$-space are suppressed.

The above derivation relies on the fact that we can replace all the quantities of interest with their ensemble average over the fast states. This is because we are observing the system on time-scales larger than the characteristic time-scale of transitions in $\nu$-space. In this way, we are ignoring the details of the latter dynamics: whether or not it drives the system out of equilibrium does not play a role here. Mathematically speaking, Eq. \eqref{Seff} does not change whether $\boldsymbol{\Phi}_i$ satisfies detailed balance ($\pi^\nu_i\phi_i^{\nu\to \mu}=\pi^\mu_i\phi_i^{\mu\to \nu},\ \forall~i$) or not.

In the following subsection, we compare the above entropy production \eqref{Seff}, derived in the framework we name CGA, with the one obtained considering the information about each single process, i.e., using SIA.

\subsection{Information on single processes (SIA)}
\label{sec:single}
Let us suppose that we are able to identify all processes acting on the system, each one due to a different coupling, and labelled by $\nu = 1,\dots, n$.

To employ the SIA, we start from the complete expression for the entropy production \eqref{S-mult}, and substitute the expression for $p^\nu_i$ given by the time-scale separation, Eqs. \eqref{expansion} and \eqref{0thorder}. Up to the zeroth order in $\epsilon$, we get the following system entropy production:
\begin{align}
\dot{S}_{sys} &=\dot S^X_{sys} + \sum_{\nu=1}^{n} \sum_{i,j} \pi_j^\nu w^\nu_{j\to i}p_j \log\dfrac{\pi_j^\nu w^\nu_{j\to i}p_j}{\pi_i^\nu w^\nu_{i\to j}p_i}+\nonumber\\
&- \sum_{\nu=1}^{n} \sum_{i,j} \pi_j^\nu w^\nu_{j\to i}p_j~ \log\dfrac{w^\nu_{j\to i}}{w^\nu_{i\to j}},
\label{ssep}
\end{align}
where second and third terms on the right hand side correspond to the total and environmental entropy production arising from the state space transitions, respectively.
Here, in order to be consistent, in the environmental contribution we split the term for each $\nu$, as we have done before in the most general case while getting Eq.~\eqref{S-mult}.

A similar formula for the entropy production is shown in Eq. \eqref{fastSFPE} for a system with continuous and discrete variables.

In Eq.~\eqref{ssep}, the entropy production associated to the transitions in the $\nu$-space is
\begin{align}
\dot S^X_{sys}&=\dfrac{1}{2\epsilon}\sum_{i,\mu,\nu}p_i(\tilde{\phi}_{i}^{\mu\to\nu} \pi^\mu_i -\tilde{\phi}_{i}^{\nu\to\mu} \pi^\nu_i )\log\dfrac{\tilde{\phi}_{i}^{\mu\to\nu}\pi^\mu_i }{\tilde{\phi}_{i}^{\nu\to\mu}\pi^\nu_i }+\nonumber\\
&- \dfrac{1}{2\epsilon}\sum_{i,\mu,\nu}p_i(\tilde{\phi}_{i}^{\mu\to\nu} \pi^\mu_i -\tilde{\phi}_{i}^{\nu\to\mu} \pi^\nu_i )\log\dfrac{\tilde{\phi}_{i}^{\mu\to\nu}}{\tilde{\phi}_{i}^{\nu\to\mu}}+\nonumber\\
&+\dfrac{1}{2}\sum_{i,\mu,\nu}(\tilde{\phi}_{i}^{\mu\to\nu} p_i^{\mu 1}-\tilde{\phi}_{i}^{\nu\to\mu} p_i^{\nu 1})\log\dfrac{\tilde{\phi}_{i}^{\mu\to\nu}\pi^\mu_i }{\tilde{\phi}_{i}^{\nu\to\mu}\pi^\nu_i }+\nonumber\\
&- \dfrac{1}{2}\sum_{i,\mu,\nu}(\tilde{\phi}_{i}^{\mu\to\nu} p_i^{\mu 1}-\tilde{\phi}_{i}^{\nu\to\mu} p_i^{\nu 1})\log\dfrac{\tilde{\phi}_{i}^{\mu\to\nu}}{\tilde{\phi}_{i}^{\nu\to\mu}}+ \nonumber\\
&+ \dfrac{1}{2}\sum_{i,\mu,\nu}(\tilde{\phi}_{i}^{\mu\to\nu} \pi^\mu_i -\tilde{\phi}_{i}^{\nu\to\mu} \pi^\nu_i )\bigg(\dfrac{p_i^{\mu 1}}{\pi^\mu_i }-\dfrac{p_i^{\nu 1}}{\pi^\nu_i }\bigg).
\label{SX}
\end{align}
The first and third terms in the above equation correspond to the total entropy production whereas the second and fourth ones correspond to the environmental entropy production. The last term appears as an extra contribution while considering the fast transitions approximation in the $\nu$-space. Thus, the entropy production given in Eq. \eqref{ssep} requires the knowledge of all above terms. 

Since now the information about the $\nu$-space is not integrated out, we have to discuss the properties of the dynamics on the fast states to proceed further. In particular, let us first consider the case in which the transition matrix governing the evolution in the $\nu$-space, $\boldsymbol{\Phi}_i$, satisfies the detailed balance [case i) Subsec. \ref{ts-d-d}]. In this case, if the label $i$ were frozen, the system would reach equilibrium in the $\nu$-space. From Eq.~\eqref{solveeq}, the detailed balance condition on $\boldsymbol{\Phi}_i$ corresponds to $\tilde{\phi}_{i}^{\mu\to\nu} \pi^\mu_i=\tilde{\phi}_{i}^{\nu\to\mu} \pi^\nu_i$ for each $i$. With this assumption, the entropy production $\dot{S}^X_{sys}$ still exhibits a correction which depends on $p_i^{\mu 1}$:
\begin{align}
\dot S^X_{sys} &= - \dfrac{1}{2}\sum_{i,\mu,\nu}(\tilde{\phi}_{i}^{\mu\to\nu} p_i^{\mu 1}-\tilde{\phi}_{i}^{\nu\to\mu} p_i^{\nu 1})\log\dfrac{\tilde{\phi}_{i}^{\mu\to\nu}}{\tilde{\phi}_{i}^{\nu\to\mu}}.
\label{correctionDB}
\end{align}
This can be seen as an extra contribution to the system entropy production which survives also when the $\nu$-space relaxes towards equilibrium (e.g. in the limit of equilibrated baths). In other words, even when the transitions taking place in the $\nu$-space are fast and detailed balanced, the (system) entropy production keeps track of first order terms in $\epsilon$ through $p_i^{\nu 1}$.

The system entropy production can be further split into total and environmental contributions, with the latter containing terms with the logarithm of ratio of transition rates. Hence, the extra contribution in Eq. \eqref{correctionDB} can be incorporated into the environmental part. This implies that the total entropy production is:
\begin{equation}
\dot{\mathcal{S}}_{tot}=\sum_{\nu=1}^{n}  \sum_{i,j} \pi_j^\nu w^\nu_{j\to i} p_j \log\dfrac{ \pi^\nu_j w^\nu_{j\to i} p_j}{\pi^\nu_i w^\nu_{i\to j} p_i}
\label{EPSIA}
\end{equation}
This expression is analogous to the one presented in \cite{esposito-CG,diffbaths} for a system in contact with multiple reservoirs. Using the log-sum inequality one derives that
\begin{equation}
\dot{\mathcal{S}}_{tot} \geq \dot S_{tot}(\tilde w,p) \label{ineq}
\end{equation}
where the right hand side  is given by Eq.~\eqref{Seff} with the definition \eqref{weff} of coarse-grained transition rates. The equality holds in Eq.~\eqref{ineq} if and only if $\pi^\nu_j w^\nu_{j\to i} p_j= c_{ij}\pi^\nu_i w^\nu_{i\to j} p_i$, where $c_{ij}$ is a constant independent of $\nu$. The trivial case $c_{ij} = 1$ corresponds to the detailed balance condition [see Eq.~\eqref{solp}], in which both sides of Eq. \eqref{ineq} are zero. However, there exist feasible solutions for $c_{ij}$ such that the system is out of equilibrium, but still both Eqs. \eqref{Seff} and \eqref{EPSIA} have the same value (see Appendix \ref{appB} for a simple example).

Note that in this formalism we find a connection between transition rates appearing in similar formulas previously derived in the literature, without explicitly using SIA, and the microscopic underlying dynamics, Eq. \eqref{MEmult}.


\subsection{State-independent fast transitions}
\label{SI-GB}
Here, we analyse how the entropy production obtained using SIA [Eqs. \eqref{ssep} and \eqref{SX}] changes when the transitions in the $\nu$-space do not depend on $i$, i.e., $\phi^{\mu\to\nu}_i \equiv \phi^{\mu\to\nu}$. This simplification leads to the conclusion that also $\pi^\nu$ are independent of the index $i$. 

Within this simple assumption, one can perform the summation over $i$-variables in \eqref{SX}. Using Eq. \eqref{normcond}: 
\begin{align}
\dot S^X_{sys}&=\dfrac{1}{2\epsilon}\sum_{\mu,\nu}(\tilde{\phi}^{\mu\to\nu} \pi^\mu-\tilde{\phi}^{\nu\to\mu} \pi^\nu)\log\dfrac{\tilde{\phi}^{\mu\to\nu}\pi^\mu}{\tilde{\phi}^{\nu\to\mu}\pi^\nu}+\nonumber\\
&-\dfrac{1}{2\epsilon}\sum_{\mu,\nu}(\tilde{\phi}^{\mu\to\nu} \pi^\mu-\tilde{\phi}^{\nu\to\mu} \pi^\nu)\log\dfrac{\tilde{\phi}^{\mu\to\nu}}{\tilde{\phi}^{\nu\to\mu}} \label{ssystiindep}.
\end{align} 
Employing also that the transition matrix $\boldsymbol{\Phi}$ satisfies detailed balance condition in the $\nu$-space, i.e., $\tilde{\phi}^{\mu\to\nu} \pi^\mu=\tilde{\phi}^{\nu\to\mu} \pi^\nu$, $\dot S^{X}_{sys}$ vanishes. Finally, up to zeroth order of $\epsilon$, from Eq. \eqref{ssep} one obtains
\begin{align}
\dot{S}_{sys}&=\sum_{\nu=1}^{n}\pi^\nu \sum_{i,j} w^\nu_{j\to i}p_j \log\dfrac{w^\nu_{j\to i}p_j}{w^\nu_{i\to j}p_i}+\nonumber\\&-\sum_{\nu=1}^{n}\pi^\nu \sum_{i,j} w^\nu_{j\to i}p_j \log\dfrac{w^\nu_{j\to i}}{w^\nu_{i\to j}}.
\label{Ssep}
\end{align}
Hence, when $\boldsymbol{\Phi}$ is detailed balanced and $i$-independent, the total entropy production corresponds to Eq.~\eqref{EPSIA} with $\pi^\nu_i \to \pi^\nu$. Notice that, in this simple case there is no correction both in system and environmental entropy production due to the first order solution in $\epsilon$.

\subsection{Broken detailed balance and time-scales}
\label{BDB}
In this section, we show the limits of applicability of the expression for the total entropy production reported in Eq. \eqref{EPSIA}, and in some previous works \cite{esposito-CG,esposito-3f,diffbaths}. To this aim, we consider the case in which the matrix $\boldsymbol{\Phi_i}$ does not satisfy detailed balance, i.e., cases ii) and iii) in Subsec. \ref{ts-d-d}.

Intuitively, the more the fast dynamics breaks detailed balance, the more it has to be faster than all other processes to not affect the quantification of non-equilibrium features, i.e., the entropy production in this context. This trade-off can indeed be quantified.


For the sake of simplicity, we restrict ourselves to an $i$-independent transition matrix $\boldsymbol{\Phi}$. From Eq. \eqref{SX}, when detailed balance is broken, some corrective terms do appear. However, in order to investigate the trade-off between characteristic time-scale and non-equilibrium stationarity, we consider the case in which detailed balance is only slightly broken. In formulas, we have:
\begin{align}
\tilde{\phi}^{\mu \to \nu} \pi^{\mu} - \tilde{\phi}^{\nu \to \mu} \pi^\nu = \xi \tilde j^{\mu \to \nu},
\label{def-xi}
\end{align}
where $\tilde j^{\mu \to \nu}$ is the scaled probability flux, and $\xi$ a small parameter quantifying the out-of-equilibrium behaviour. On a fairly general level, a breakage of detailed balance stems from the injection of energy in the system. It may be manifested in several forms, as, for example, an imposed thermal \cite{diss-sel} or chemical gradient \cite{chemo, granick}, chemostatted concentrations \cite{rao}, or a constant light irradiation \cite{photoacids}. Due to Eqs. \eqref{solveeq} and \eqref{0thorder} we must have that $\sum_{\nu=1}^n \tilde j^{\mu \to \nu}=0$. An interesting perspective might arise from framing $\xi$ in the context of linear response theory for Markovian systems \cite{lrm}. We leave this for future discussions.


From Eqs.~\eqref{ssep} and \eqref{ssystiindep}, the total entropy production in this condition can be identified as, (up to the zeroth order in $\epsilon$)
\begin{eqnarray}
\dot{\mathcal{S}}_{tot} &=& \sum_{\nu=1}^{n} \pi^\nu \sum_{i,j} w^\nu_{j\to i} p_j \log\dfrac{w^\nu_{j\to i} p_j}{w^\nu_{i\to j} p_i} + \nonumber \\
&+& \dfrac{1}{2\epsilon}\sum_{\mu,\nu}(\tilde{\phi}^{\mu\to\nu} \pi^\mu-\tilde{\phi}^{\nu\to\mu} \pi^\nu)\log\dfrac{\tilde{\phi}^{\mu\to\nu}\pi^\mu}{\tilde{\phi}^{\nu\to\mu}\pi^\nu}.
\end{eqnarray}
Expanding the last term in the above equation up to the leading order in $\xi$, we get:
\begin{eqnarray}
\dot{\mathcal{S}}_{tot} &=& \sum_{\nu=1}^{n} \pi^\nu \sum_{i,j} w^\nu_{j\to i} p_j \log\dfrac{w^\nu_{j\to i} p_j}{w^\nu_{i\to j} p_i} + \nonumber \\
&\;& + ~\frac{\xi^2}{2 \epsilon} \sum_{\mu,\nu} \frac{ \tilde j_{\mu \to \nu}^2}{\tilde{\phi}^{\nu \to \mu} \pi^{\nu}}
\label{xieps-1}
\end{eqnarray}
giving a quantification of the interplay between broken detailed balance and time-scale separation, encoded in the ratio of two expansion parameters, $\xi^2/\epsilon$. Note that the term multiplying this pre-factor has the same form of the total entropy production in $\nu$-space \cite{busielloENT}.

Similar corrections to the entropy production due to slightly broken detailed balance condition for a system governed by Eq. \eqref{FPEmult} can be seen in Eq. \eqref{xieps}.

An imperative remark is that, given the expression of the total entropy production in Eqs. \eqref{xieps-1} and \eqref{xieps}, the presence of non-equilibrium conditions in the $\nu$-space, encoded in $\xi$, could prevent the possibility to perform a consistent time-scale separation on the system dynamics. In fact, even if the rates $\phi^{\mu \to \nu}$ are much faster than all the others, non-equilibrium effects could lead to non-vanishing corrective terms of order $\epsilon^{-1}$ in the entropy production. Naively speaking, there exist situations in which different time-scales are entangled regardless of the level of description. This is in accordance to what has been shown in \cite{busielloENT,jstat,celani}: the dissipation keeps track of microscopic degrees of freedom which would have been ignored describing the system \textit{ab initio} through a coarse-grained (i.e., approximate, zeroth-order) dynamics.

In the following, we present some illustrative examples in which the typical scales $\xi$ and $\epsilon$ assume physical meanings. We aim at hinting at the working conditions under which a biological system can be effectively described performing a time-scale separation, while keeping information about each single process, i.e., using Eq. \eqref{EPSIA} for the total entropy production with multiple coupling.

\section{Examples}
\label{examples}
In this section, we present some simple cases in which our framework can be applied. We show under which conditions the entropy production is affected by the interplay between detailed balance and the fastest time-scale, translating the condition derived in this paper in terms of physical quantities.

\subsection{Molecular motors}

At first, we consider a molecular motor moving along a one-dimensional ring, having a potential landscape described by $U(x)$. A schematic diagram is shown in Fig.~\ref{fig:system}B. This model describes all families of motor proteins: kynesins and dynesins moving along tubulin filaments, and myosins along actin filaments \cite{parmeggiani1999energy,julicher}.

Here, we refer to the standard framework presented in \cite{qian}. The molecular motor can be in different states (tracks), each one following its own diffusion equation. Moreover, the system can change state by consumption of the fuel, e.g. hydrolyzing ATP. As a consequence of the interplay between different processes, a linear directive motion is induced in the system. For simplicity, we assume that there are only two configurations. Then, the equation governing the dynamics of the molecular motor is 
\begin{align}
\dfrac{\partial \vec{P}(x,t)}{\partial t}=-\dfrac{\partial \vec{J}}{\partial x}+\Phi \vec{P}(x,t),
\label{m-e-mm}
\end{align}  
where $\vec{P}=(P_1(x,t),P_2(x,t))^\top$, $\vec{J}=(J_1(x,t),J_2(x,t))^\top$, where $J_i(x,t)=-\mu_i[-k_B T\partial_x P_i(x,t)+\{-\partial_x U_i(x)+f_{ext}\}P_i(x,t)]$, and  
\begin{align*}
\boldsymbol{\Phi}(x)=\begin{pmatrix}
-w_1(x) && w_2(x)\\
w_1(x) && -w_2(x)
\end{pmatrix}
\end{align*} 
is the matrix capturing the transitions of the molecular motor among the tracks. Notice that its sum over each column is equal to zero to ensure probability conservation.

Using the above model, we can find the entropy production as given in Eq. \eqref{SdotFPE}. It is equal to the one shown in Eqs. (22) and (23) in Ref.~\cite{qian} in the stationary state. In the following, we further perform a time-scale separation analysis, assuming that the transitions between different states are faster than the diffusion along each track. A similar analysis on a general setup is shown in Appendix \ref{ts-cd-model}. Now, expanding the solution of the master equation \eqref{m-e-mm} as in Eq. \eqref{expansion}
\begin{align}\label{soln-eqn}
\vec{P}(x,t) \approx \vec{P}^{(0)}(x,t)+\epsilon \vec{P}^{(1)}(x,t).
\end{align}
and plugging Eq.~\eqref{soln-eqn} in the master equation \eqref{m-e-mm}, we find that $\vec{P}^{(0)}(x,t)=\vec{\Pi}(x) \mathcal{P}(x,t)$, where $\vec{\Pi}(x)=[\Pi^{(1)}(x),\Pi^{(2)}(x)]^\top$ is the stationary solution of the fast dynamics, governed by the transition matrix $\boldsymbol{\Phi}(x)$, and $\mathcal{P}(x,t)$ is the effective probability density function (see Appendix~\ref{ts-cd-model}). From the model given above, we can exactly find $\vec{\Pi}(x)=[{w_2}(x),{w_1}(x)]^\top/[{w_1}(x)+{w_2}(x)]$.

An important remark can be made from the above calculations: when only two tracks are present, $\boldsymbol{\Phi}(x)$ is always detailed balanced in the fast time-scale approximation, and the only correction to the entropy production may arise from the terms of first order in $\epsilon$ [for e.g. see Eq. \eqref{fastSFPE}]. Finally, if the transition rates across the tracks are also independent of the spatial variable, the contribution due to first order correction also disappears and the entropy production only depends on the driving along the tracks. In the following, we consider a simple case when the detailed balance condition in the fast space can also be violated.

\subsubsection*{Multiple configurations and breakage of detailed balance}
Herein, we modify the problem discussed above, admitting the existence of several (more than two) internal configurations among which the particle can switch. In this case the detailed balance can be broken in the internal space of tracks, even in the limit of fast transitions. 

We consider the most general case in which detailed balance is slightly broken in the fast space. The latter is characterized by the spatial dependent matrix $\boldsymbol{\Phi}(x)$. We are aiming at understanding how non-equilibrium features entangle to fast time-scales, leading to extra contributions to the entropy production, and under which conditions on physical parameters such a contribution is not negligible.

In particular, in the limit of fast internal transitions (among tracks), the following term appears in the entropy production:
\begin{eqnarray}
\frac{\xi^2}{2 \epsilon} \sum_{\mu,\nu} \frac{\tilde j_{\mu \to \nu}^2}{\tilde{\phi}^{\nu \to \mu}(x) \Pi^{\nu}(x)}
\label{xiepsexample}
\end{eqnarray}
which contains information both on fast processes and non-equilibrium behavior. Here, $\xi$ is the magnitude of the mechanism keeping the system away from equilibrium. Then, we can see that if the product of the square of the strength of the deviation from equilibrium condition, as measured by $\xi$ and its own time scale $1/\epsilon$ remains finite, and non zero, i.e., $\xi^2/\epsilon \sim \mathcal{O}(1)$, the total entropy production is affected by an additional non-vanishing quantity, even in the limit of infinitely fast transitions and very slight out-of-equilibrium conditions. It is important to notice that such a contribution is due to the microscopic fluxes among all possible configurations which are present in the system, as evidenced by the term $\tilde j_{\mu \to \nu}$ in the equation above.

As a toy model, let us consider a system composed by three tracks, each one with its own energy landscape, $U_\nu(x)$. Molecules can move on each of them, according to three different diffusion equations. Moreover, they can also pass from one track to the other with the following transition rates:
\begin{align}
\phi^{\mu \to \nu} = \phi^{\nu \to \mu} e^{\Delta U_{\mu,\nu}(x)/(k_B T)}   \quad  \text{for all}\quad \mu,\nu,
\label{db-10}
\end{align}
where $T$ the temperature of the environment, $k_B$ the Boltzmann constant, and $\Delta U_{\mu,\nu}(x) = U_\mu(x) - U_\nu(x)$. For simplicity, let us imagine that one particular transition rate (from $\mu^*$ to $\nu^*$) is modified by the presence of a chemical potential difference, $\Delta c$:
\begin{equation}
\phi^{\mu^* \to \nu^*}= \phi^{\nu^* \to \mu^*} e^{(\Delta U_{\mu^*,\nu^*}(x) - \Delta c)/(k_B T)},    
\end{equation}
When $\Delta c_{\mu,\nu}/(k_B T)$ is small, we can expand the (only) flux flowing in the track-space, finding that 
${ \pi^\mu \tilde\phi^{\mu \to \nu}- \pi^\nu \tilde\phi^{\nu \to \mu}} \propto \Delta c/(k_B T)$, for all $\mu$ and $\nu$. Hence, $\Delta c/(k_B T)$ plays the role of $\xi$, quantifying how much the system is out of equilibrium. Thus, we have:
\begin{equation}
    \xi = \frac{\Delta c}{k_B T} \qquad \text{and}\qquad \epsilon = \frac{\tilde{\phi}_{\mu \to \nu}}{\phi_{\mu \to \nu}},
\end{equation}
where the second of the previous equation is just the definition of $\epsilon$ (see Sec. \ref{ts-d-d}).

If the motion along each single track, independently, would reach equilibrium, the system would not produce entropy based solely on the motion along the tracks. Mathematically, this corresponds to
\begin{equation}
\int dx \frac{(J^{\nu}(x,t))^2}{D^\nu(x) P(x,t)} = 0,
\end{equation}
for the $\nu = 1,2,3$.

On the other hand, if the motion among tracks is not at equilibrium, because of the chemical potential difference, detailed balance is slightly broken. In the limit of fast transitions among tracks, considering all the typical scales in play, the correction to the total entropy production becomes non-negligible when the following scaling holds:
\begin{equation}
\Delta c \sqrt{\frac{\phi_{\mu \to \nu}}{\tilde{\phi}_{\mu \to \nu}}} \sim k_B T.
\end{equation}
Hence, even if the first contribution to the entropy production in Eq. \eqref{xiepsexample} is zero, the second one becomes relevant when the chemical potential differences become comparable to the available thermal energy for each transition.

A similar dynamically and thermodynamically consistent coarse-graining procedure for molecular motors in the presence of probe particles is discussed in \cite{probe}.

\subsection{Three-state chemical reaction network in a temperature gradient}




Here, we present another example which is a slight generalization of the one extensively studied in \cite{diss-sel}, inspecting the possibility to select high-energy metastable states at stationarity via non-equilibrium processes and energy dissipation.

The system consists of three chemical states: $A$, $B$, and $C$, and the transitions among them are defined by
\begin{align}
A \xleftrightarrow[\kappa_{B \to A}]{\kappa_{A \to B}} B \xleftrightarrow[\kappa_{C \to B}]{\kappa_{B \to C}} C \xleftrightarrow[\kappa_{A \to C}]{\kappa_{C \to A}} A
\end{align}
The system can also diffuse between spatially separated baths at different temperatures.

Following the original article \cite{diss-sel}, if all transition rates satisfy Arrenhius' relations, the detailed balance in the chemical (internal) space is always respected. Going further, let us consider that the rate from $B$ to $C$ is enhanced by a quantity $\Delta e/(k_B T)$, because, for example, of the presence of a catalytic molecule, so that the system can attain a non-equilibrium steady state in the chemical space. In the limit of fast reactions, and close to equilibrium conditions, we can identify two small parameters:
\begin{gather}
    \xi = \frac{\Delta e}{k_B T(x)} \;\;\;\;\; \epsilon = \frac{\tilde{\kappa}_{X \to Y}}{\kappa_{X \to Y}}
\end{gather}
where $X,Y = A,B,C$ and $\tilde{\kappa}_{X \to Y}$ is the rescaled reaction rate from $X$ to $Y$ leading to the identification of the small parameter $\epsilon$. The working condition here employed, e.g. fast reactions, serves only as an example. Indeed, as long as one subspace supports faster transitions, our framework can be applied. However, there are experimental settings in which the stirring of solutions at different temperatures can be externally controlled, modifying, in turn, the effective diffusion coefficient \cite{chiral}.

A quantity that naturally appears in the context of dissipation-driven phenomena is $L_k = \sqrt{D/\kappa_{X \to Y}}$ \cite{diss-sel}. This is the characteristic length at which the system can absorb and dissipate energy through diffusive cycles. It is possible to write the condition $\xi^2/\epsilon \sim \mathcal{O}(1)$, letting this quantity appear, as follows:
\begin{equation}\label{46}
\frac{\Delta e}{L_k} \sim \frac{k_B T(x)}{\sqrt{D/\tilde{\kappa}_{X \to Y}}}
\end{equation}
Notice that, while $L_k$ is a characteristic length, $\sqrt{D/\tilde{\kappa}_{X \to Y}}$ contains information only about the typical scale of the diffusion, since $\tilde{\kappa}_{X \to Y}$ is defined as the rate rescaled by its magnitude, $\epsilon$.

This means that, when the non-equilibrium energy density over the typical dissipation length , the left hand side of Eq. \eqref{46}, is at least of the same order of the available energy density over the typical diffusive length, right hand side  of Eq. \eqref{46}, the entropy production has a non-vanishing contribution stemming from the interplay between non-equilibrium conditions and the fastest dynamics. In other words, $\dot{S}_{tot}$ is affected by the presence of diffusive cycles dissipating energy via fast chemical fluxes, represented by $\tilde{j}_{\mu \to \nu}$ in Eq. \eqref{xieps-1}, which is the discrete counterpart of Eq. \eqref{xiepsexample}.

\subsection{Catalytic enzymes}

As another biologically-inspired example, let us consider the case of an enzyme $E$, which can catalyze the transformation of a substrate $S$ into a product $P$. Moreover, it can bind/unbind both to $S$ and $P$ with different rates, forming complexes. In the simple, yet quite common setting in which the enzyme is much bigger than the substrate \cite{sen,granick}, its diffusion coefficient can be considered similar to the one of the complexes. The reaction network characterizing the system can be schematized as follows:
\begin{gather}
E \xleftrightarrow[k_{S \to E}]{k_{E \to S}[S]} E + S, \nonumber \\
E \xleftrightarrow[k_{P \to E}]{k_{E \to P}[P]} E + P, \\
E + S \xleftrightarrow[k_{P \to S}]{k_{S \to P}} E + P. \nonumber
\end{gather}
In the above equation, the transition rate above the arrow is intended to pertain to the left-to-right transition. Here, $E$ indicates the free enzyme in solution, with $S$ and $P$ floating around with concentration $[S]$ and $[P]$, respectively. The states $E+S$ and $E+P$ are bound states (complexes with $S$ and $P$). In many experimental settings, $[S]$ and $[P]$ are chemostatted or externally controlled, maintaining the system in a non-equilibrium steady state, at a given energy cost. A quantification of the latter is given by the deviation of the ratio between the two concentrations from the equilibrium value:
\begin{equation}
r = \frac{[S]}{[P]} = e^{\Delta S_m} \frac{[S]^{\rm eq}}{[P]^{\rm eq}} = e^{\Delta S_m} r^{\rm eq}
\end{equation}
where $\Delta S_m$ is the entropy change in the environment. Hence, the quantity $r$ quantifies how far the system is from being at equilibrium. One notable case is when $[S]$ and $[P]$ corresponds to ATP and ADP concentrations, respectively, and $r$ accounts for the available energy in the system \cite{exc-hyd,assenza}. In the latter case, the enzyme catalyzes ATP hydrolysis.

As for the previous examples, let us analyse the situation in which the enzyme feels a substrate gradient, $[S(x)]$ \cite{granick}. In this case, it has been shown that it is a good approximation to consider chemical interactions to be much faster than diffusion \cite{chemo}. If we are also in close to equilibrium conditions, $\xi = r/r^{\rm eq} \approx 1$, the typical scaling allowing for a non-negligible additional contribution to the entropy production, i.e., the second term in Eq. \eqref{xiepsexample}, is:
\begin{equation}
\frac{r}{L_S} \sim \frac{r^{\rm eq}}{\sqrt{D/\tilde{k}_{X \to Y}}}
\end{equation}
Here, $\epsilon = \tilde{k}_{X \to Y}/k_{X \to Y}$ as for the previous case, with the subscripts $X$ and $Y$ indicating, in general, any two possible states of the system. Analogously, $L_S = \sqrt{D/k_{X \to Y}}$ is the energy absorption-dissipation characteristic length. Also in this case, we can write this condition in terms of energy density, noting that if the available energy over the dissipative length-scale $L_S$ is of the same order with respect to its equilibrium value in a purely diffusive system, the entropy production is affected by microscopic fluxes in the fast space [$\tilde j^{\mu \to \nu}$ in Eq. \eqref{xiepsexample}].

\subsection{Multi-state particles in contact with switching baths}

Finally, we study the case of a multi-state particle whose transitions are triggered by the coupling to $n$ thermal baths.

In the literature \cite{esposito-CG,diffbaths}, the entropy production has been derived to be always equal to Eq. \eqref{xieps-1} with $\xi = 0$. Here, we have shown that this is just an approximation of the most general case. In fact, it implicitly assumes that the dynamics in the bath space is faster than all other processes, hence employing what we called SIA. Moreover, other necessary conditions to obtain the entropy production as in \cite{esposito-CG,diffbaths} are that the dynamics in the bath space is detailed balanced and space-independent. More specifically, a particular case satisfying all these assumptions is
\begin{gather}
\phi^{\mu \to \nu} = \phi^{\nu \to \mu} \Rightarrow \pi^\nu = \frac{1}{n}
\end{gather}
The equation above implies that the effective transition rates derived from the SIA are trivially proportional to the original rates of the slow process.

However, it is important to note that, when the dynamics in the reservoir space, governed by the transition matrix $\boldsymbol{\Phi}(x)$, is fast, but not detailed balanced, the total entropy production has to be corrected. In particular, when $\boldsymbol{\Phi}(x)$ does not depend on $x$, we have that, even if the system does not produce entropy according to the slow dynamics only, i.e., Eq. \eqref{solp} at stationarity satisfies
\begin{equation}
    \sum_\nu \pi^{\nu} w^\nu_{i \to j} p_i - \sum_\nu \pi^{\nu} w^\nu_{j \to i} p_j = 0 \;\;\;\;\; \forall~i,j
\end{equation}
the total entropy production still does not vanish, because of the non-zero contribution proportional to $\xi^2/\epsilon$. This extra term takes into account fluxes among reservoirs, $\tilde j^{\mu \to \nu}$. Here, the physical meaning of the scaling relation, $\xi^2/\epsilon \sim \mathcal{O}(1)$, has to be determined on a single case basis. In general, it is worth noting that our proposed formula for the entropy production can be markedly different from the previously derived one.

\section{Conclusions}
\label{conc}

Non-equilibrium features are sensibly affected by coarse-graining procedures. This general statement has a long-standing tradition, and it has been proved in many different contexts \cite{busielloENT,jstat,celani}. However, some approximations exist and are usually employed to describe non-equilibrium systems without unnecessary details \cite{esposito-CG}.

In this paper, we dealt with systems in the presence of multiple coupling. In other words, we have studied the interplay between different classes of transitions, generated by different processes, acting on the same system. Most of biological systems belong to this category (e.g.  molecular motors \cite{julicher}, enzymes \cite{sen}, chemical reaction networks \cite{rao,diss-sel}).

Two widely used approximations can be applied in the presence of multiple coupling: SIA and CGA, both extensively discussed throughout the paper. Here, as a first step, we have explicitly derived them from a general framework both for discrete and continuous state-spaces.

Both SIA and CGA rely on the assumption that some processes are much faster than all the others. The CGA erases all information about the latter while the SIA applies a weaker coarse-graining, and some details about the fast dynamics is retained. Well-known formulas previously obtained in the literature can be reconstructed within the SIA. As a further step, we have identified the physical conditions under which our general framework leads to some extra contributions to the entropy production, with respect to these formulas. These latter terms are, in fact, signatures of an intrinsic non-equilibrium condition, and as such, they can be substantially affected by any kind of coarse-graining procedure.

Indeed, it is possible to determine a scaling relation between the amount of breakage of detailed-balance, named $\xi$, and the characteristic time-scale of the faster processes, named $\epsilon$, such that the entropy production will differ from the one known in the literature. Intuitively speaking, even if one process is very fast without leading to an equilibrium state in its sub-space, particularly strong microscopic fluxes can be entangled with the slow process, producing a non-vanishing \textit{extra} entropy production at the macroscopic level.

In the last part of the paper we have presented some simple, yet instructive, systems in which the scaling relation between $\xi$ and $\epsilon$ can be translated into a relation among physical quantities. These can serve both to unveil the role of detailed balance and time-scales in some pedagogical examples, and to capture the main ingredients (and their interplay) that allow simplified theoretical analyses of chemical (or biological) minimal models. However, since our approach is rather general, \textit{mutatis mutandis}, it is amenable of application even in more complex settings.

\section*{Acknowledgments}

D. Gupta and A. Maritan acknowledge the support from University of Padova through “Excellence Project 2018” of the Cariparo foundation.

\appendix
\section{Entropy production with transitions among discrete $\nu$-space and diffusive dynamics}
\label{ep-c-d}
In the following, we consider a system with a continuous $i$-space (let us call it $x$-space) and a discrete $\nu$-space (e.g.,  chemical states, reservoirs). The case in which both of them are continuous is a straightforward generalization. This system can also be described within the framework of Master Equation as Eq.~\eqref{FPEmult} \cite{gardiner}.

For sake of simplicity, we consider a system that moves along a one-dimensional ring, whose evolution is governed by the following overdamped Langevin equation:
\begin{align}
\dot{x}=\dfrac{F^\nu(x,t)}{\gamma^\nu(x)}+\sqrt{2D^\nu(x)} \eta(t),
\end{align}
where $F^\nu(x,t)=-\partial_x U^\nu(x,t)+f^\nu(x,t)$ is the external force acting on the system which can be decomposed into the force arise from the confining potential $U^\nu(x,t)$ and a non-conservative external force $f^\nu(x,t)$. Herein, $\gamma^\nu(x)$ and $D^\nu(x)$, respectively, are the space-dependent dissipation and diffusion coefficient. Notice that the superscript labels the $\nu$-space. In the above equation, $\eta(t)$ is a Gaussian white noise with mean zero and unit variance: $\langle\eta(t)\rangle=0$ and $\langle\eta(t)\eta(t')\rangle=\delta(t-t')$, where the angular brackets indicate the averaging over the noise distribution. 

The probability of the system to be at position $x$ and in state $\nu$ evolves according to Eq. \eqref{FPEmult}, where 
\begin{align}
J^\nu(x,t)=\dfrac{F^{\nu}(x,t)P^\nu(x,t)}{\gamma^\nu(x)}-\dfrac{\partial (D^\nu(x) P^\nu(x,t))}{\partial x},
\end{align}
is the probability current.

In this case, the (average) entropy of the system is 
\begin{align}
S_{sys}=-\sum_\mu \int dx\ P^\mu(x,t) \log P^\mu(x,t). 
\end{align}

Differentiating with respect to time and using the normalization condition $\sum_\nu \int dx\ P^\nu(x,t)=1$, the system entropy production becomes 
\begin{align}
\dot{S}_{sys}&=-\sum_\mu \int dx\ \dfrac{\partial P^\mu(x,t)}{\partial t}\log P^\mu(x,t) = \nonumber\\
&=\sum_\mu \int dx\  \bigg[\dfrac{\partial J^\mu(x,t)}{\partial x}\log P^\mu(x,t)\bigg]+\nonumber\\
&\; + \dfrac{1}{2}\sum_{\mu,\nu} \int dx\   \bigg[\phi^{\mu\to \nu}(x)P^\mu(x,t)+\nonumber\\
&\; -\phi^{\nu\to \mu}(x)P^\nu(x,t)\bigg]\log\dfrac{P^\mu(x,t)}{P^\nu(x,t)},
\end{align}
where, going from first equality to the second one, we have used Eq.~\eqref{FPEmult}. Integrating by parts the first term on the right-hand side and substituting the definition of the current $J^\nu(x,t)$, we get
\begin{align}
&\dot{S}_{sys}=\overbrace{\sum_\mu \int dx\ \dfrac{{J^\mu(x,t)}^2}{D^\mu(x)P^\mu(x,t)}}^{\dot S_{tot}}+\nonumber\\
&+\overbrace{\sum_{\mu,\nu}\int dx\ \phi^{\mu\to \nu}(x)P^\mu(x,t)~\log\dfrac{\phi^{\mu\to \nu}(x)P^\mu(x,t)}{\phi^{\nu\to \mu}(x)P^\nu(x,t)}}^{\dot S^X_{tot}}\nonumber\\
&-\bigg[\overbrace{\sum_\mu \int dx\ \bigg[\dfrac{A^\mu(x,t) J^\mu(x,t)}{D^\mu(x,t)}-J^\mu(x,t) \dfrac{\partial }{\partial x} \log D^\mu(x)\bigg]}^{\dot S_{env}}+\nonumber\\
&-\overbrace{\sum_{\mu,\nu}\int dx\ \phi^{\mu\to \nu}(x)P^\mu(x,t)\log\dfrac{P^\mu(x,t)}{P^\nu(x,t)}}^{\dot S^X_{env}}\bigg],
\label{SdotFPE}
\end{align}
where we have defined~$A^\nu(x,t)=F^\nu(x,t)/\gamma^\nu(x)$. 
In the above equation, we have imposed the periodic boundary conditions on the probability current $J^\mu(x,t)$ for each $\mu$. The splitting here shown is analogous to the one presented in Eq. \eqref{S-mult}.

When the external force and the potential are time-independent, the system \textit{asymptotically} reaches the steady state. At stationarity, the left-hand side of the above equation \eqref{SdotFPE} vanishes, and the right-hand side is satisfied by $P_{ss}^\nu(x)$, where the subscript $ss$ indicates the non-equilibrium stationary state.

\subsection{Time-scale separation on the dynamics}
\label{ts-cd-model}
As shown in section \ref{ts-d-d}, here we can also consider that the transition occurring in the $\nu$-space are faster with respect to all other possible transitions, i.e., $\phi^{\mu\to \nu}(x)=\tilde{\phi}^{\mu\to \nu}(x)/\epsilon$, where $\epsilon$ is the characteristic time-scale. Therefore, we get
\begin{align}
\dfrac{\partial P^\nu(x,t)}{\partial t}&=-\dfrac{\partial J^\nu(x,t)}{\partial x}~+\nonumber\\
&+\dfrac{1}{\epsilon}\sum_{\mu=1}^{n} [\tilde{\phi}^{\mu\to \nu}(x)P^\mu(x,t)-\tilde{\phi}^{\nu\to \mu}(x)P^\nu(x,t)].
\label{s-d-me}
\end{align}
We assume the solution of the above equation (up to first order in $\epsilon$) as
\begin{equation}
P^\nu(x,t) \approx P^{\nu 0}(x,t)+\epsilon~P^{\nu 1}(x,t).
\end{equation}
Substituting the above solution in Eq. \eqref{s-d-me} and comparing the terms of similar orders in $\epsilon$ yields 
\begin{align}
\dfrac{dP^{\nu 0}(x,t)}{dt}&= -\dfrac{\partial }{\partial x}J^\nu(x,t)\bigg|_{P^{\nu}(x,t)\to P^{\nu 0}(x,t)}+\nonumber\\
&+\sum_{\mu}[\tilde{\phi}^{\mu\to \nu}(x)P^{\mu 1}(x,t)-\tilde{\phi}^{\nu\to \mu}(x)P^{\nu 1}(x,t)],\label{s-d-pn0}\\
\dfrac{dP^{\nu 1}(x,t)}{dt}&=-\dfrac{\partial }{\partial x}J^\nu(x,t)\bigg|_{P^{\nu}(x,t)\to P^{\nu 1}(x,t)},\\
\nonumber\\
0&=\sum_{\mu}[\tilde{\phi}^{\mu\to \nu}(x)P^{\mu 0}(x,t)-\tilde{\phi}^{\nu\to \mu}(x)P^{\nu 0}(x,t)]\label{pi-eqn}.
\end{align}  
The above equation \eqref{pi-eqn} implies that the quantity $P^{\mu 0}(x,t)$ reaches stationary state in the $\nu$-space. Therefore, $P^{\mu 0}(x)=\Pi^{\nu}(x)\mathcal{P}(x,t)$. Notice that $\Pi^\nu(x)$ is the space-dependent stationary distribution with respect to transition rates in the $\nu$-space, $\phi^{\mu \to \nu}(x)$, and it is normalized as $\sum_\nu \Pi^\nu(x)=1$.

Summing over the discrete state in Eq. \eqref{s-d-pn0}, we get 
\begin{align}\label{eqn-cg-2}
\dfrac{\partial \mathcal{P}(x,t)}{\partial t}&=-\dfrac{\partial }{\partial x}\overbrace{\bigg(\tilde{A}(x,t)\mathcal{P}(x,t)-\dfrac{\partial (\tilde{D}(x) \mathcal{P}(x,t))}{\partial x}\bigg)}^{\tilde{J}(x,t)},
\end{align}
where  $\mathcal{P}(x,t) = \sum_\nu \Pi_\nu(x) \mathcal{P}(x,t)$, due to the normalization of $\Pi_\nu(x)$, is defined as an effective probability distribution. In analogy to what we have done in Sec. \ref{sec:one}, we also introduce an effective drift, $\tilde{A} = \sum_{\nu} \Pi^\nu(x) A^\nu$, and an effective diffusion coefficient, $\tilde{D} = \sum_{\nu} \Pi^\nu(x) D^\nu$, which are nothing but the ensemble average of $A^\nu(x)$ and $D^\nu(x)$ over the $\nu$-space.

\subsection{Coarse-grained variables (CGA)}
\label{cg-process}
Retracing all the steps extensively discussed  in Sec. \ref{sec-cga}, here we consider the case where we cannot distinguish each single fast state $\nu$, and the only accessible information is about coarse-grained quantities. Then, we write the Master equation for the coarse-grained probability density function by summing Eq.~\eqref{FPEmult} over the discrete variables as
\begin{equation}
\dfrac{\partial \mathcal{P}(x,t)}{\partial t} =-\sum_\nu\dfrac{\partial J_\nu(x,t) }{\partial x}
\label{dynglob}
\end{equation}
where $\mathcal{P}(x,t)=\sum_\nu P_\nu(x,t)$, i.e.,~the coarse-grained probability distribution. We can define the entropy of the system in this case as following
\begin{align}
S_{sys}(\mathcal{P})=-\int dx\ \mathcal{P}(x,t) \log \mathcal{P}(x,t). 
\end{align}
Differentiating the above equation with respect to time, we obtain the system entropy production as
\begin{align}
\dot S_{sys}(\mathcal{P})&=-\int dx\ \dfrac{\partial \mathcal{P}(x,t)}{\partial t} \log \mathcal{P}(x,t)
\label{entprodglob}
\end{align}
We made use only of the accessible coarse-grained quantities to define the system entropy.

Employing the fast time-scale approximation for the $\nu$-space, i.e.,  $P^\nu(x,t)=\Pi^\nu(x) \mathcal{P}(x,t) + \epsilon P^{\nu 1}(x,t)$, we rewrite Eq.~\eqref{entprodglob} using Eq.~\eqref{eqn-cg-2} as a function of $\tilde{A}$ and $\tilde{D}$. After some simple manipulation, we have:
\begin{align}
\label{cg-c-d-sys}
\dot S_{sys}(\mathcal{P})&=\int dx\ \frac{\partial \tilde{J}(x,t)}{\partial x} \log \mathcal{P}(x,t) = \nonumber \\
&= - \int dx\ \frac{\tilde{J}(x,t)}{\mathcal{P}(x,t)} \frac{\partial \mathcal{P}(x,t)}{\partial x} = \nonumber \\
&= \overbrace{\int dx\ \dfrac{\tilde{J}^2(x,t)}{\tilde{D}(x)\mathcal{P}(x,t)}}^{\dot S_{tot}}+\nonumber\\
&-\overbrace{\int dx\ \bigg[\dfrac{\tilde{A}(x,t) \tilde{J}(x,t)}{\tilde{D}(x,t)}-\tilde{J}(x,t) \dfrac{\partial }{\partial x} \log \tilde{D}(x)\bigg]}^{\dot S_{env}}
\end{align}
As for the discrete state space (see Sec. \ref{sec-cga}), we do not need further assumptions on the dynamics of the $\nu$-space. Indeed, if they are fast, and we cannot discriminate among them, this is enough to have a total entropy production which depends just on the slow states. However, drift and diffusion coefficient have to be substituted with their ensemble average over the fast states. As a consequence, environmental entropy production is identified only as a function of accessible variables in this approximation.

\subsection{Information on single processes (SIA)}
\label{fast-bath}
In the following, we employ the time-scale separation on the entropy production given in Eq.~\eqref{SdotFPE}, according to the procedure characterizing the SIA. We have:
\begin{widetext}
\begin{align}
\dot S_{sys} &=\sum_\nu \int dx\ \bigg(\dfrac{(\bar{J}^\nu(x,t))^2}{\bar{D}^\nu(x)\mathcal{P}(x,t)}-\dfrac{\bar{J}^\nu(x,t) \bar{A}^\nu(x,t)}{\bar{D}^\nu(x)}+\bar{J}^\nu(x,t)\dfrac{\partial}{\partial x}\log \bar{D}^\nu(x)\bigg)+\dfrac{1}{2}\sum_{\mu, \nu} \int dx\ \bigg[\dfrac{1}{\epsilon} \mathcal{P}(x,t) \big[\tilde{\phi}^{\mu\to\nu}(x)\Pi^\mu(x)+\nonumber\\
& -\tilde{\phi}^{\nu\to\mu}(x)\Pi^\nu(x)\big]\log \dfrac{\Pi^\mu(x)}{\Pi^\nu(x)}+\big[\tilde{\phi}^{\mu\to\nu}(x)P^{\mu 1}(x,t)-\tilde{\phi}^{\nu\to\mu}(x)P^{\nu 1}(x,t)\big]\log \dfrac{\Pi^\mu(x)}{\Pi^\nu(x)}+\nonumber\\
& +\big[\tilde{\phi}^{\mu\to\nu}(x)\Pi^\mu(x)-\tilde{\phi}^{\nu\to\mu}(x)\Pi^\nu(x)\big]\bigg(\dfrac{P^{\mu 1}(x,t)}{\Pi^\mu(x)}-\dfrac{P^{\nu 1}(x,t)}{\Pi^\nu(x)}\bigg) \bigg],
\label{fastSFPE}
\end{align}
\end{widetext}
where the integrals refer to to the motion in the continuum space, whereas the summation are performed over the states belonging to the discrete $\nu$-space. In the above equation, we introduce the following notation: $\bar{Z}^\nu:=Z^\nu \Pi^\nu(x)$.

Eq. \eqref{fastSFPE} is the analogous of Eqs. \eqref{ssep} and \eqref{SX} of the main text.

\subsection{Detailed-balance and time-scales}

Simplified formulas for the total entropy production can be recovered starting from Eq. \eqref{fastSFPE}. In particular, when the matrix $\boldsymbol{\Phi}(x)$ is detailed balanced, i.e., $\tilde{\phi}^{\mu\to\nu}(x)\Pi^\mu(x)=\tilde{\phi}^{\nu\to\mu}(x)\Pi^\nu(x)$ for each $x$, the first and third terms in the second summation on the right hand side of Eq. \eqref{fastSFPE} becomes zero, and the remaining terms contribute to the entropy production. In the simplest case in which the dynamics in the $\nu$-space is also independent of space, noticing that $\int dx~P^{\nu 1}(x,t) = 0$, also the second term in the second summation on the right hand side vanishes.

Conversely, if the detailed balance in the $\nu$-space is slightly broken, non-equilibrium effects can entangle fast and slow time-scales. This provides an expression for the entropy production which is not consistent with an adiabatic elimination of fast variables $\textit{a-priori}$ in the dynamics. In other words, employing the time-scale separation before or after the estimation of the entropy production does not lead to the same result: the former case corresponds to the CGA, while the latter is the SIA. In complete analogy to what has been shown in the main text, now we have:
\begin{eqnarray}
\dot{\mathcal{S}}_{tot} &=& \sum_{\nu=1}^{n} \int dx \frac{J^2_{\nu}(x,t)}{D^\nu(x) P^\nu(x,t)} + \nonumber \\
&\;& + ~\frac{\xi^2}{2 \epsilon} \sum_{\mu,\nu} \frac{j_{\mu \to \nu}^2}{\tilde{\phi}^{\nu \to \mu} \Pi^{\nu}(x)},
\label{xieps}
\end{eqnarray}
assuming that $\boldsymbol{\Phi}$ does not depend on $x$, where $\xi$ is the small parameter  as introduced in Eq. \eqref{def-xi}.
\bigskip
\section{An example of equivalence of SIA- and CGA-entropy production in a 3-state system in a non-equilibrium steady state}
\label{appB}
In this section, we discuss the condition under which the entropy production in SIA and CGA becomes equal. Clearly, from Eq. \eqref{ineq}, the equality holds when 
\begin{eqnarray}
\pi^\nu_i w^\nu_{i \to j} p_i = c_{ij} \pi^\nu_j w^\nu_{j \to i} p_j
\end{eqnarray}
Plugging this condition into Eq. \eqref{solp}, we have:
\begin{equation}
\frac{dp_i}{dt} = \sum_j \sum_\nu \pi^\nu_i w_{i \to j} p_i (c_{ij} - 1) = K_{i} p_i,
\end{equation}
where we have defined $K_i= \sum_j \sum_\nu \pi^\nu_i w_{i \to j} (c_{ij} - 1)$.

Since $p_i$ is a probability, the only feasible solutions must satisfy the condition $K_i = 0$. This can trivially happens when $c_{ij} = 1$, $\forall~i,j$, i.e., at equilibrium. However, here we show a simple example in which $K_i = 0$ even if the system is in a non-equilibrium steady state.

Consider a $3$-state model with the following effective transition rates:
\begin{gather}
\tilde{w}_{1 \to 2} = 2 \;\;\; \tilde{w}_{2 \to 3} = 3 \;\;\; \tilde{w}_{3 \to 1} = 4 \nonumber \\
\tilde{w}_{2 \to 1} = 1 \;\;\; \tilde{w}_{3 \to 2} = 1 \;\;\; \tilde{w}_{1 \to 3} = 3
\end{gather}
The product of transition rates in the clockwise direction is larger than that in anti-clockwise direction. Hence, the system sustains a non-zero probability current at the stationary state [see Eq. \eqref{solp}].
 
Notice that $\tilde{w}_{i \to j} = \sum_\nu \pi^\nu_i w_{i \to j}$, so they can be obtained in several ways starting from the microscopic rates. However, $K_i$ depends only on effective quantities. The condition $K_i = 0$ is fulfilled when:
\begin{equation}
c_{1 \to 2} = \frac{13}{34} \;\;\; c_{2 \to 3} = \frac{6}{13} \;\;\; c_{1 \to 3} = \frac{24}{17}
\end{equation}
with $c_{ij} = c_{ji}^{-1}$. Hence, the two approximations (SIA) and (CGA) lead to the same value of the entropy production even if the system is in a non-equilibrium stationary state.

A similar result, in which the effect of the coarse-graining vanishes for particular choices of the currents is also presented in \cite{busielloENT}.


\begin{thebibliography}{}
\bibitem{prigogine}
Prigogine, I. and Nicolis, G., 1971. Biological order, structure and instabilities. Quarterly reviews of biophysics, 4(2-3), pp.107-148.

\bibitem{rao}
Rao, R. and Esposito, M., 2016. Nonequilibrium thermodynamics of chemical reaction networks: wisdom from stochastic thermodynamics. Physical Review X, 6(4), p.041064.

\bibitem{semenov}
Semenov, S.N., Kraft, L.J., Ainla, A., Zhao, M., Baghbanzadeh, M., Campbell, V.E., Kang, K., Fox, J.M. and Whitesides, G.M., 2016. Autocatalytic, bistable, oscillatory networks of biologically relevant organic reactions. Nature, 537(7622), pp.656-660.




\bibitem{schn}
Schnakenberg, J., 1976. Network theory of microscopic and macroscopic behavior of master equation systems. Reviews of Modern physics, 48(4), p.571.


\bibitem{diffbaths}
Tomé, T. and de Oliveira, M.J., 2012. Entropy production in nonequilibrium systems at stationary states. Physical review letters, 108(2), p.020601.

\bibitem{diss-sel}
Busiello, D.M., Liang, S.L. and Rios, P.D.L., 2019. Dissipation-driven selection in non-equilibrium chemical networks. arXiv preprint arXiv:1912.04642.

\bibitem{fickslaw}
Asllani, M., Busiello, D. M., Carletti T., Fanelli D. and Planchon, G., 2014, Turing patterns in multiplex networks. Physical Review E, 90(4), p.042814

\bibitem{gardiner}
Gardiner, C., 2009. Stochastic methods (Vol. 4). Berlin: Springer.


\bibitem{Schmiedl_2007}
Schmiedl, T. and Seifert, U., 2007. Efficiency at maximum power: An analytically solvable model for stochastic heat engines. EPL (Europhysics Letters), 81(2), p.20003.

\bibitem{PhysRevLett.99.230602}
Filliger, R. and Reimann, P., 2007. Brownian gyrator: A minimal heat engine on the nanoscale. Physical review letters, 99(23), p.230602.


\bibitem{exc-hyd}
Goloubinoff, P., Sassi, A.S., Fauvet, B., Barducci, A. and De Los Rios, P., 2018. Chaperones convert the energy from ATP into the nonequilibrium stabilization of native proteins. Nature chemical biology, 14(4), pp.388-395.

\bibitem{Wachtel_2018}
Wachtel, A., Rao, R. and Esposito, M., 2018. Thermodynamically consistent coarse graining of biocatalysts beyond Michaelis–Menten. New Journal of Physics, 20(4), p.042002.


\bibitem{julicher}
Jülicher, F., Ajdari, A. and Prost, J., 1997. Modeling molecular motors. Reviews of Modern Physics, 69(4), p.1269.

\bibitem{parmeggiani1999energy}
Parmeggiani, A., Jülicher, F., Ajdari, A. and Prost, J., 1999. Energy transduction of isothermal ratchets: Generic aspects and specific examples close to and far from equilibrium. Physical Review E, 60(2), p.2127.

\bibitem{sen}
Mohajerani, F., Zhao, X., Somasundar, A., Velegol, D. and Sen, A., 2018. A theory of enzyme chemotaxis: from experiments to modeling. Biochemistry, 57(43), pp.6256-6263.


\bibitem{van1992stochastic}
Van Kampen, N.G., 1992. Stochastic processes in physics and chemistry (Vol. 1). Elsevier.


\bibitem{jstat}
Busiello, D.M. and Maritan, A., 2019. Entropy production in master equations and Fokker–Planck equations: facing the coarse-graining and recovering the information loss. Journal of Statistical Mechanics: Theory and Experiment, 2019(10), p.104013.


\bibitem{busielloENT}
Busiello, D.M., Hidalgo, J. and Maritan, A., 2019. Entropy production for coarse-grained dynamics. New Journal of Physics, 21(7), p.073004.


\bibitem{seifFT}
Seifert, U., 2012. Stochastic thermodynamics, fluctuation theorems and molecular machines. Reports on progress in physics, 75(12), p.126001.



\bibitem{esposito-CG}
Esposito, M., 2012. Stochastic thermodynamics under coarse graining. Physical Review E, 85(4), p.041125.


\bibitem{gillespie}
Gillespie, D.T., 1991. Markov processes: an introduction for physical scientists. Elsevier.


\bibitem{busiello-raz}
Busiello DM, Jarzynski C, Raz O. Similarities and differences between non-equilibrium steady states and time-periodic driving in diffusive systems. New Journal of Physics. 2018 Sep 13;20(9):093015.


\bibitem{seifEP}
Seifert, U., 2005. Entropy production along a stochastic trajectory and an integral fluctuation theorem. Physical review letters, 95(4), p.040602.



\bibitem{pigo}
Pigolotti, S., Neri, I., Roldán, É. and Jülicher, F., 2017. Generic properties of stochastic entropy production. Physical review letters, 119(14), p.140604.

\bibitem{unidirectional}
Busiello, D.M., Gupta, D. and Maritan, A., 2020. Entropy production in systems with unidirectional transitions. Physical Review Research, 2(2), p.023011.



\bibitem{barato}
Barato, A.C. and Seifert, U., 2016. Cost and precision of Brownian clocks. Physical Review X, 6(4), p.041053.





\bibitem{gupta2020thermodynamic}
Gupta, D. and Maritan, A., 2020. Thermodynamic uncertainty relations in a linear system. The European Physical Journal B, 93(2), pp.1-8.


\bibitem{friedman2020thermodynamic}
Friedman, H.M., Agarwalla, B.K., Shein-Lumbroso, O., Tal, O. and Segal, D., 2020. Thermodynamic uncertainty relation in atomic-scale quantum conductors. Physical Review B, 101(19), p.195423.



\bibitem{gingrichtur}
Gingrich, T.R., Horowitz, J.M., Perunov, N. and England, J.L., 2016. Dissipation bounds all steady-state current fluctuations. Physical review letters, 116(12), p.120601.


\bibitem{pietztur}
Pietzonka, P., Barato, A.C. and Seifert, U., 2016. Universal bounds on current fluctuations. Physical Review E, 93(5), p.052145.

\bibitem{falascotur}
Falasco, G., Esposito, M. and Delvenne, J.C., 2020. Unifying thermodynamic uncertainty relations. New Journal of Physics.



\bibitem{baratotur}
Barato, A.C. and Seifert, U., 2015. Thermodynamic uncertainty relation for biomolecular processes. Physical review letters, 114(15), p.158101.



\bibitem{hyper}
Busiello, D.M. and Pigolotti, S., 2019. Hyperaccurate currents in stochastic thermodynamics. Physical Review E, 100(6), p.060102.



\bibitem{van2020entropy}
Van Vu, T. and Hasegawa, Y., 2020. Entropy production estimation with optimal current. Physical Review E, 101(4), p.042138.


\bibitem{horo}
Li, J., Horowitz, J.M., Gingrich, T.R. and Fakhri, N., 2019. Quantifying dissipation using fluctuating currents. Nature communications, 10(1), pp.1-9.

\bibitem{guptainfer}
Manikandan, S.K., Gupta, D. and Krishnamurthy, S., 2020. Inferring entropy production from short experiments. Physical Review Letters, 124(12), p.120603.



\bibitem{otsubo2020estimating}
Otsubo, S., Ito, S., Dechant, A. and Sagawa, T., 2020. Estimating entropy production by machine learning of short-time fluctuating currents. Physical Review E, 101(6), p.062106.


\bibitem{metab}
Niebel B., Leupold S., and Heinemann M., 2019. An upper limit on Gibbs energy dissipation governs cellular metabolism. Nature Metabolism 1(1), 125-132.


\bibitem{qian}
Qian, M., Zhang, X., Wilson, R.J. and Feng, J., 2008. Efficiency of Brownian motors in terms of entropy production rate. EPL (Europhysics Letters), 84(1), p.10014.

\bibitem{SKM}
Manikandan, S.K., Dabelow, L., Eichhorn, R. and Krishnamurthy, S., 2019. Efficiency fluctuations in microscopic machines. Physical review letters, 122(14), p.140601.



\bibitem{jarz-raz}
Raz, O., Subaşı, Y. and Jarzynski, C., 2016. Mimicking nonequilibrium steady states with time-periodic driving. Physical Review X, 6(2), p.021022.


\bibitem{mehl}
Mehl, J., Lander, B., Bechinger, C., Blickle, V. and Seifert, U., 2012. Role of hidden slow degrees of freedom in the fluctuation theorem. Physical review letters, 108(22), p.220601.
\bibitem{kawa}
Kawaguchi, K. and Nakayama, Y., 2013. Fluctuation theorem for hidden entropy production. Physical Review E, 88(2), p.022147.
\bibitem{uhl}
Uhl, M., Pietzonka, P. and Seifert, U., 2018. Fluctuations of apparent entropy production in networks with hidden slow degrees of freedom. Journal of Statistical Mechanics: Theory and Experiment, 2018(2), p.023203.

\bibitem{eff-th}
Polettini, M. and Esposito, M., 2017. Effective thermodynamics for a marginal observer. Physical review letters, 119(24), p.240601.


\bibitem{pep-1}
Gupta, D. and Sabhapandit, S., 2016. Fluctuation theorem for entropy production of a partial system in the weak-coupling limit. EPL (Europhysics Letters), 115(6), p.60003.
\bibitem{pep-2}
Gupta, D. and Sabhapandit, S., 2020. Entropy production for partially observed harmonic systems. Journal of Statistical Mechanics: Theory and Experiment, 2020(1), p.013204.
\bibitem{pep-3}
Gupta, D. and Sabhapandit, S., 2018. Partial entropy production in heat transport. Journal of Statistical Mechanics: Theory and Experiment, 2018(6), p.063203.



\bibitem{esposito-3f}
Esposito, M. and Van den Broeck, C., 2010. Three faces of the second law. I. Master equation formulation. Physical Review E, 82(1), p.011143.



\bibitem{celani}
Celani, A., Bo, S., Eichhorn, R. and Aurell, E., 2012. Anomalous thermodynamics at the microscale. Physical review letters, 109(26), p.260603.

\bibitem{granick}
Jee, A.-Y., Dutta, Sandipan, Cho Y.-K., Tlusty T., and Granick S., 2018. Enzyme leaps fuel antichemotaxis. PNAS 115(1), 14-18.


\bibitem{chemo}
Busiello D. M., De Los Rios P., and Piazza F., 2020. Nonequilibrium theory for enzyme chemotaxis and enhanced diffusion. arXiv preprint arXiv:2007.14242.



\bibitem{photoacids}
Berton C., Busiello D. M., Zamuner S., Solari E., Scopelliti R., Fadaei-Tirani F., Severin K., and Pezzato C., 2020. Thermodynamics and kinetics of protonated merocyanine photoacids in water. Chemical Science, 11(32), 8457-8468.



\bibitem{lrm}
Andrieux, D. and Gaspard, P., 2007. A fluctuation theorem for currents and non-linear response coefficients. Journal of Statistical Mechanics: Theory and Experiment, 2007(02), p.P02006.



\bibitem{probe}
Zimmermann, E., and Seifert, U., 2015. Effective rates from thermodynamically consistent coarse-graining of models for molecular motors with probe particles. Physical Review E, 91, 022709.



\bibitem{chiral}
Viedma, C. and Cintas, P., 2011. Homochirality beyond grinding: deracemizing chiral crystals by temperature gradient under boiling. Chemical Communications, 47, 12786-12788.


\bibitem{assenza}
Assenza, S., Sassi, A.S., Kellner, R., Schuler, B. and Alessandro, B., 2019. Efficient conversion of chemical energy into mechanical work by Hsp70 chaperones. eLife, 8, e48491.




















\end{thebibliography}

\end{document}